\documentstyle[12pt]{article}
\newcounter{bean}
\newcounter{pea}

\newcommand{\ret}{\hfill\break\noindent}
\def\Pslash{\hbox{P\kern-.7em\raise.2ex\hbox{/}}}

\begin{document}

\begin{flushright} Preprint USM-TH-67 \end{flushright}
\medskip

\begin{center} \Large 3-dimensional rules for finite-temperature loops
\end{center}
\vspace*{.15in}
\normalsize
\begin{center}
C. Dib\footnote{cdib@fis.utfsm.cl}
, O. Espinosa
~and I. Schmidt
\end{center}

\vspace*{.15in}
\begin{center}
Department of Physics \\ Universidad T\'{e}cnica Federico Santa Mar\'\i a\\
Casilla 110-V \\ Valpara\'\i so, Chile  
\end{center}

\begin{abstract}
We present simple diagrammatic rules to write down Euclidean $n$-point
functions at finite temperature directly in terms of 3-dimensional momentum 
integrals, without ever performing a single Matsubara sum. 
The rules can be understood as describing the interaction of the external 
particles with those of the thermal bath. 
\end{abstract}

\vspace{.25in}

\newpage
\noindent{\bf Introduction:}
In many applications of perturbative field theory at finite temperature it 
is necessary to compute higher-order corrections to Green functions, i.e., 
loop diagrams.
In one approach \cite{Landsman and van Weert}, {\em the real-time formalism}, 
one computes Green functions in real time, i.e., thermal averages of 
time-ordered products of Heisenberg field operators. 
The perturbative evaluation of these Green functions entails a duplication of 
the degrees of freedom (through the introduction of ghost fields), which 
either increases the multiplicity of diagrams or forces the use of 
matrix-valued propagators.
In the {\em Euclidean} or {\em imaginary-time formalism}, on the other hand, 
one considers the so-called temperature Green functions, i.e., thermal averages
of $\tau$-ordered Heisenberg field operators in imaginary time $t=-i\tau$, 
with $0\leq\tau\leq 1/T$.
The Fourier transforms of these temperature Green functions are defined only 
for discrete values of the (Euclidean) energy, $P^0=\omega_n$ 
($n=0,\pm1,\ldots$), with $\omega_n=(2n)\pi T$ for bosons and 
$\omega_n=(2n+1)\pi T$ for fermions, the so-called Matsubara frequencies.
In this formalism there is no need for ghost fields and the Feynman rules are 
very similar in form to those of zero-temperature Euclidean field theory, 
except for the fact that now each loop in the diagram is associated with an 
infinite sum over Matsubara frequencies, instead of an integration over a 
continuous energy variable.
Although these sums can be computed in a number of ways, usually in a 
systematic fashion, these computations can become quite tedious for higher 
loop diagrams. 
In one procedure, the so-called {\em Saclay method} \cite{Braaten+Pisarski}, 
one uses a Fourier integral representation of the Euclidean propagator in 
terms of a variable $\tau$ conjugate to the discrete Matsubara loop energy 
$K^0$. 
In this form the energy sums are easily done, giving rise to delta functions 
in the $\tau$-variables, which, in a graph with $I$ internal lines and $L$ 
loops, can be used to eliminate $L$ of the $I$ $\tau$-integrals. 
What remains are $\tau$-integrals of exponential functions, each of them 
giving rise to a denominator linear in energies. 
Although straightforward, in practice the calculations usually become 
cumbersome, and a simple form for the final result can only be obtained after 
due use of some of the identities satisfied by the Bose-Einstein and 
Fermi-Dirac thermal occupation numbers \cite{Braaten+Pisarski-identities}.
Another way to perform the Matsubara sums is the standard contour integration 
method, in which the summation over the energy is represented as a contour 
integral with an appropiately chosen integrand.
In reference \cite{Guerin} this method was used, along with a decomposition 
of the Euclidean propagator into partial fractions, to compute general 
$N$-point graphs. Certain rules are given that allow to write down the 
general answer directly in terms of the external energies.
In both methods, the result for a general amputated scalar graph with $N+1$ 
vertices ($N\geq 1$), $I$ internal lines, and external 4-momenta
$P_l=(P_{l}^0, {\bf p}_l)$, has the form
\begin{equation}
\frac{(-g)^{N+1}}S\int
[\prod\limits_{i=1}^{I}\frac{d^3k_i}{\left( 2\pi \right) ^32E_i}%
\prod\limits_{V=1}^N (2\pi )^3\delta^{(3)} ({\bf k}_V)]
\;D(P_{l}^0,E_i), 
\label{eq:saclay-euclidean}
\end{equation}
\ret
where $g$ represents the coupling constant, $S$ is the symmetry factor of the 
graph, the $E_i=({\bf k}_i^2+m_i^2)^{1/2}$ are the energies of the internal 
lines (carrying spatial momenta ${\bf k}_i$), and ${\bf k}_V$ denotes the 
total momentum entering vertex $V$. 
The delta functions ensure conservation of spatial momentum at each vertex, 
so that the integration measure reduces essentially to an integration over 
the 3-momenta of the $L=I-N$ independent loops. 
The integrand in (\ref{eq:saclay-euclidean}) is of the form 
$\sum_\alpha N_\alpha/R_\alpha$, a sum of rational expressions, where each 
numerator $N_\alpha$ is a polynomial of degree $L$ in the occupation numbers 
$n_i\equiv n(E_i)=(\exp(E_i/T)-1)^{-1}$, and each denominator $R_\alpha$ is 
a product of exactly $N$ factors, all linear in the energies $iP_{l}^0$ 
and $E_i$.

In this letter we present simple diagrammatic rules to write down an explicit 
expression for the integrand $D(P_{l}^0,E_i)$, appearing in 
(\ref{eq:saclay-euclidean}), for any finite-temperature Euclidean Feynman 
graph, thereby completely avoiding the need to perform Matsubara sums.
Although similar rules have been given in Ref. \cite{Guerin}, we believe that 
ours are simpler and explicitly incorporate the known fact that finite 
temperature amplitudes can be related to zero-temperature forward scattering 
amplitudes in a thermal bath \cite{Frenkel+Taylor}.
The rules can be formulated in two alternative ways, in terms of two different 
types of diagrams.
In one case, the diagrams are closely related to those encountered in 
old-fashioned zero-temperature time-ordered perturbation theory\cite{Schweber},
and lead to a result that strongly resembles in form the one obtained by using 
standard Euclidean Feynman rules, once the Matsubara sums have been performed.
A second, simpler form of the rules can be obtained from the preceding one by 
regrouping similar diagrams into standard Feynman diagrams, thus reducing 
drastically the number of diagrams to be considered.
In both approaches, the diagrams include cut (or ``snipped'') lines that 
represent the interaction of the external particles with the thermal 
bath.

We shall first consider diagrams for scalar particles. 
In this case the integrand $D$ depends on the {\em spatial} momenta only 
through the energies of the internal lines $E_i$. 
We will indicate at the end the necessary modifications for fermions. 

\medskip
\noindent{\bf The Rules:}
In our first approach, which is analogous to the old-fashioned time-ordered 
diagrammatic expansion, the Feynman graph for the case at hand is given by 
expression (\ref{eq:saclay-euclidean}), with the integrand $D(P_{l}^0,E_i)$ 
computed according to the following rules:

\begin{list}
{\alph{bean}.}{\usecounter{bean}}
\item 
For each external line, characterized by a real Euclidean 4-vector 
$(P_{l}^0, {\bf p}_l)$, define the {\em energy} of the line as $iP_{l}^0$. 
For each internal line define its {\em energy} as $E_i=({\bf k}_i^2+m_i^2)^{1/2}$, 
where ${\bf k}_i$ is the 3-momentum carried by the line and $m_i$ is the mass 
of the propagating particle.
\item 
Define a {\em direction of time} or {\em energy flow} (which we shall take
conventionally from left to right) and consider all possible orderings of the 
vertices along this direction (see, e.g., Figs.~1.a and 1.b. For a graph with 
$N+1$ vertices there will be $(N+1)!$ such orderings).
\item 
For each time-ordered graph generated in (b) consider, in addition to 
itself, all possible {\em connected} graphs that can be obtained by
snipping any number of internal lines.
Each line that is snipped becomes a pair of legs we shall call {\em thermal 
legs}.
Attach a cross to their ends to distinguish them from the original external 
lines of the graph. 
Both legs of a given pair inherit the energy $E_i$ of the internal 
line that originated them. 
However, one leg must be oriented as {\em incoming}
with energy $E_i$ and the other as {\em outgoing} with energy $E_i$. 
Both possible orientations have to be considered, each one generating a 
different diagram (see, e.g., Figs.~1.c and 1.d).
\item 
For each graph in (c), define its total incoming energy, $E_{inc}$, as 
the sum of all incoming external energies plus the energies of all incoming 
thermal legs that join the diagram {\em before} their outgoing partner 
(e.g.~Figs.~1.d and 1.f). 
Thermal leg pairs that satisfy this property shall be referred to as 
{\em external}, and those that do not as {\em internal} (e.g.~Figs.~1.c 
and 1.e). 
Then, associate to this graph an expression equal to the product of the 
following factors:
\begin{list}
{\arabic{pea}.}{\usecounter{pea}}
\item 
draw a full vertical division (a ``cut'') between each pair of consecutive 
time-ordered vertices (there are $N$ such cuts in a graph with $N+1$ vertices);
for each cut include a factor
\begin{equation}
\frac {1}{E_{inc}-E_{cut}} 
\label{eq:energy-denominator}
\end{equation}
where $E_{cut}$ is the total energy of the intermediate state associated with 
the cut, defined as the sum of the energies of all the lines that cross the 
cut in question (as in zero-temperature time-ordered perturbation theory),
plus the energies of all {\em internal} thermal pairs whose originating 
internal line would have crossed the cut.
\item 
include a thermal occupation factor $n_i\equiv n(E_i)$ for each thermal
pair (of energy $E_i$) in the diagram (if any).
\item 
include an overall factor of $(-1)^N$, where $N+1$ is the number of vertices.
\end{list}
\end{list}
The integrand $D(P_{l}^0,E_i)$ is the sum of the expressions computed 
according to rule (d), over all the graphs in (c).

Since the temperature $T$ enters only through the thermal occupation factors 
$n_i$ described in rule d.2, it is evident that the full result separates 
explicitly into a zero-temperature and a temperature-dependent part, the 
former being obtained solely from the ``un-snipped'' diagrams, since the thermal 
occupation factors vanish for $T=0$. 
This is a known property of the Euclidean finite-temperature result, once the 
Matsubara sums have been dealt with.
Rule (c), in conjunction with rule (d.2), embodies the known fact that any 
term contributing to the temperature-dependent part of a $L$-loop graph 
contains at most $L$ factors of the thermal occupation numbers. 
In fact, in a $L$-loop graph one can snip at most $L$ internal propagators if 
the resulting graph is to remain connected. 

As it stands, the diagrammatic method gets out of hand for diagrams with more 
than a few vertices, due to the factorial growth in the number of time-ordered 
diagrams (see rule (b)).
We show below that the time-ordered graphs can be regrouped into standard 
``covariant'' Feynman diagrams, thereby simplifying the computation 
considerably. These diagrams are similar to standard $T=0$ Feynman diagrams,
except for the inclusion of thermal leg pairs.

A general proof of the above diagrammatic rules will be presented elsewhere
\cite{In-preparation}. 
We have verified their correctness in a few simple cases by hand and in 
various nontrivial cases with {\em Mathematica}.
In the remainder of this letter we would like just to illustrate how the rules 
work in a few examples.
We shall leave the analysis of the general case for Ref. \cite{In-preparation}.

\medskip
\noindent{\bf Example 1:}
We consider first the lowest order correction to the inverse propagator in the 
scalar $g\phi^3$ theory (see Fig.~1), which we will call 
$L_2(P)$, with $P=(P^0, {\bf p})$. 
A straightforward application of the Saclay method gives 
\begin{equation}
L_2(P)=\frac{g^2}{2}\int \frac{d^3k}{\left( 2\pi \right) ^3}\frac 1{2E_12E_2}
\;D_2(P^0,E_1,E_2), 
\label{eq:L2-euclidean}
\end{equation}
\ret
with
\vbox{
\begin{eqnarray}
\lefteqn{D_2(P^0,E_1,E_2)=-\frac{1+n_1+n_2}{iP^0-E_1-E_2}
+\frac{n_1-n_2}{iP^0-E_1+E_2}%
-\frac{n_1-n_2}{iP^0+E_1-E_2} } \nonumber\\
& &
\hspace{8.5truecm}+\frac{1+n_1+n_2}{iP^0+E_1+E_2}, \qquad
\label{eq:D2-euclidean}
\end{eqnarray}
}
\ret
where $E_1=({\bf k}^2+m^2)^{1/2}$ and $E_2=[({\bf p-k})^2+m^2]^{1/2}$ are the 
energies of the internal lines.

In our approach, each one of the ten terms in equation (\ref{eq:D2-euclidean})
corresponds to one particular snipped, time-ordered diagram\footnote{This is true
in this particular simple case. 
In general, there is no such simple correspondence between the Saclay method 
and our diagrams.}.
The zero-temperature contribution is obtained from diagrams 1.a and 1.b. 
These give
\begin{equation}
D_2(P^0,E_1,E_2)\Big\vert_{T=0}=-\frac 1{iP^0-E_1-E_2}+\frac
1{iP^0+E_1+E_2} 
\label{eq:D2-T=0}
\end{equation}
\ret
One can readily verify that the usual $T=0$ Feynman integral of this 1-loop
correction can be put in the general form 
(\ref{eq:saclay-euclidean}), with $D_2$ given in the equation above, by 
performing explicitly the $k_0$-integral by contour integration.
In the time-ordered formalism, the relative minus sign between both terms 
in (\ref{eq:D2-T=0}) has its origin in the energy of the intermediate state in 
diagram (2b), namely $E_{cut}=2iP^0+E_1+E_2$.

The finite temperature piece of $D_2$, according to our rules, is obtained 
from the graphs 1.c to 1.f (and the corresponding ones obtained by interchanging 
the labels of lines 1 and 2).
Their contribution to $D_2$ is computed according to rule (d), giving:
\begin{eqnarray}
\lefteqn{-n_1\Biggl[\frac{1}{iP^0-E_1-E_2}+\frac{1}{iP^0+E_1-E_2}%
-\frac{1}{iP^0+E_1+E_2} } \nonumber\\
& &
\hspace{7.5truecm}-\frac{1}{iP^0-E_1+E_2}\Biggr], \qquad
\label{eq:D2-finite-temperature}
\end{eqnarray}
\ret
which fully agrees with the Saclay result (\ref{eq:D2-euclidean}).
The diagrams 1.c to 1.f have an obvious physical interpretation as extra,
thermal contributions to the $T=0$ diagrams 1.a and 1.b, where one 
(or, in general, more than one) of the particles participating in the process 
is taken out from and put back into the thermal bath, with probability $n(E)$.

\medskip
\noindent{\bf Example 2:}
As a less trivial example, consider the 2-loop graph in the $\lambda\phi^4$ 
theory, shown in Fig.~2.a. 
As in the previous case, there are two possible time-orderings of the 
vertices, depicted in Figs.~2.a and 2.b. 
The $T=0$ contribution to the integrand $D$ is obtained with our 
rules directly from these graphs, and is equal to:
\begin{equation}
-\frac 1{iP^0-E_1-E_2-E_3}+\frac 1{iP^0+E_1+E_2+E_3} 
\label{eq:2-loop-T=0}
\end{equation}
\ret
The finite temperature contribution is obtained by snipping internal lines of
graphs 2.a and 2.b. 
As in the previous example, one has terms that are linear in the $n_i$'s, 
corresponding to cutting only one internal line. 
Two particular contributions are shown in Figs.~2.c and 2.d.
Graph 2.c is calculated with $E_{inc}=iP^0+E_1$ and $E_{cut}=E_2+E_3$ to give 
$-n_1/(iP^0+E_1-E_2-E_3)$. 
Similarly, graph 2.d has $E_{inc}=iP^0$ and $E_{cut}=2iP^0+E_1+E_2+E_3$, and 
gives $+n_1/(iP^0+E_1+E_2+E_3)$.
A new type of contribution appears in this 2-loop graph: terms bilinear in the 
$n_i$'s, obtained by cutting two internal lines. 
Again, two particular contributions are shown in Figs.~2.e and 2.f. 
For graph 2.e, $E_{inc}=iP^0+E_1+E_3$ and $E_{cut}=E_2$, which gives 
$-n_1 n_3/(iP^0+E_1-E_2+E_3)$. 
For graph 2.f, $E_{inc}=iP^0+E_3$ and $E_{cut}=2iP^0+E_1+E_2$, which gives 
$+n_1 n_3/(iP^0+E_1+E_2-E_3)$. 
The full finite-temperature result is obtained from the sum of all possible 
snipped diagrams.

\medskip
\noindent{\bf Example 3:}
As a last example we consider the one-loop contribution to the 3-point 
function in the $g\phi^3$ theory, shown in Fig.~3. 
The new ingredient here is the presence of many time-orderings and more than 
one intermediate state. 
There are 6 possible vertex orderings. For the $T=0$ part, two of these 
orderings are shown in Figs.~3.a and 3.b, and their contribution to $D$ is:
\begin{eqnarray}
\lefteqn{ \frac 1{(iP_a^0-E_1-E_2)(iP_a^0+iP_b^0-E_1-E_3)} }\nonumber\\
& &
\hspace{3truecm}
+~\frac 1{(iP_a^0+iP_b^0+E_1+E_3)(iP_a^0+E_1+E_2)} 
\label{eq:triangle-T=0}
\end{eqnarray}
\ret
The finite temperature part comes from snipping the $T=0$ graphs. From
graphs 3.a and 3.b we thus obtain, amongst others, graphs 3.c and 3.d, 
whose contribution to $D$ is:
\begin{eqnarray}
\lefteqn{ \frac {n_2}{(iP_a^0-E_1+E_2)(iP_a^0+iP_b^0-E_1-E_3)} }\nonumber\\
& &
\hspace{3truecm}
+~\frac {n_2}{(iP_a^0+iP_b^0+E_1+E_3)(iP_a^0+E_1+E_2)} 
\label{eq:triangle-finite-T}
\end{eqnarray}
\ret
Again, to obtain the full result one has to consider all possible orderings of 
the vertices and all possible snips of internal lines that keep the graph 
connected.

In the first example given above, it is easy to realize that the graphs in 
Figs.~1.c to 1.f can be combined in pairs to form ``thermally snipped'' 
{\em Feynman} diagrams. 
For instance, the first and fourth terms inside the square bracket in
(\ref{eq:D2-finite-temperature}) can be combined into the expression
\begin{equation}
2E_2\frac 1{(iP^0-E_1)^2-E_2^2},
\label{eq:combine-terms}
\end{equation}
which, apart from the factor $2E_2$ and the overall minus sign in
(\ref{eq:D2-finite-temperature}), is precisely the Euclidean
propagator in the t-channel diagram for the {\em zero-temperature} forward 
scattering process $\phi(P^0,{\bf p})+\phi(-iE_1, {\bf k})\rightarrow
\phi(P^0,{\bf p})+\phi(-iE_1, {\bf k})$, shown in Fig.~4.a.
The factor $2E_2$ cancels out a similar factor in the integration 
measure in (\ref{eq:saclay-euclidean}), so that the contributions from 
graphs 1.c and 1.f add up to 
\begin{equation}
\int \frac{d^3k}{2E_1}n_1 {\cal T}(P^0,-iE_1),
\end{equation}
where ${\cal T}(P^0,-iE_1)$ denotes the t-channel diagram in Fig.~4.a.
Denoting likewise the s-channel diagram in Fig.~4.b by ${\cal S}(P^0,-iE_1)$, 
we see 
that $L_2(P^0, {\bf p})$ can be expressed purely in terms of 
standard Feynman diagrams as:
\begin{eqnarray}
L_2(P^0, {\bf p})&&=L_2(P^0, {\bf p})\Big\vert_{T=0}+ \nonumber\\
&&\int \frac{d^3k}{2E_1}n_1 \left[{\cal T}(P^0,-iE_1)+{\cal S}(P^0,-iE_1)\right]
+ (1\leftrightarrow 2).
\end{eqnarray}

It is not difficult to realize that the ability to regroup a set of our 
time-ordered diagrams into a standard Feynman diagram is quite general. 
Consider the sum of all possible time orderings of a given diagram, with a 
fixed set of oriented thermal legs. Alternatively, consider the same set of 
time-ordered diagrams, but disregarding the thermal character of its external 
lines: this is precisely the set of time-ordered graphs in zero-temperature 
theory that add up to a single, fully covariant, Feynman diagram with the 
given set of external lines. 
Then the rules above, apart from the product of thermal factors (d.2) 
(which is the same for all the diagrams in the set), provide an alternative 
way for computing this very same zero-temperature Feynman diagram. 
In fact, it is easy to see that rule (d.1) gives the same contribution, 
regardless of whether we consider some of the thermal legs as internal or all 
of them as external.
In our simple example above, for instance, equation (\ref{eq:combine-terms}) 
corresponds to the regrouping of graphs 1.c and 1.f. 
Notice that the thermal legs attach in identical manners to the vertices in 
both graphs. 

\medskip
\noindent{\bf Rules for a simpler approach:}
There is therefore an alternative simpler way of computing a given Euclidean
finite-temperature graph:
 
\begin{list}
{\alph{bean}.}{\usecounter{bean}}
\item Generate all possible {\em connected} graphs that can be obtained from
the given graph by snipping any number of internal lines, and attach crosses to
the ends of the snipped lines (the thermal legs).
For any given thermal pair, one of the legs has to be incoming and the 
other one outgoing, both carrying the same momentum and energy 
$(-iE_i, {\bf k}_i)$.
Both possible orientations have to be considered, each one generating a 
different diagram.
\item Compute the graphs in (a) using standard Euclidean Feynman rules as in
zero-temperature and integrate each graph over the spatial momenta of its 
thermal pairs, with weight factor $n_i/2E_i$ for each pair.
\end{list}
The given Euclidean graph is equal to the sum of the contributions above plus 
its zero-temperature value.

\medskip
\noindent{\bf Final Comments:}
The modifications to our rules for the fermionic case are rather simple. 
First, there is the sacred $-1$ factor for each fermion loop. 
Second, for each fermionic (snipped or not) line of spatial momentum {\bf p} 
and mass $m_f$, the numerator gets a factor $\Pslash_s -m_f$, where 
$P_s=(-siE_p, {\bf p})$, $s=\pm1$, and where the Euclidean gamma matrices 
$\bar{\gamma}_\mu$ (given by $\{\bar{\gamma}_\mu,\bar{\gamma}_\nu\}=
-2\delta_{\mu\nu}$) must be used.
The variable $s$ takes on the value $+1$ if the fermionic line is oriented 
in the forward time direction and $-1$ otherwise. 
Finally, the thermal factor associated with a snipped fermionic line is 
$-\tilde{n}(E_k)$, where $\tilde{n}(E)=(\exp(E/T)+1)^{-1}$ is the Fermi-Dirac 
thermal occupation number.

We should point out that our method, although useful in explicit calculations,
may not be convenient for renormalization group analysis. In T=0 Feynman diagrams,
low order corrections to self-energies and vertices are computed only once; 
the results can then be inserted into higher order skeletons, which makes 
the renormalization group simple to implement. This same principle works for 
both the Matsubara method (before continuation) and the real-time method with 
doubled components. However, this feature may not work here.

As a final remark, we should point out a possible extension of our method to 
the real-time formalism. 
If one considers all Euclidean external energies $P^0_l$ as imaginary, such 
that $iP^0_l = E_l$ becomes real (i.e. an analytic continuation), includes an 
$i\epsilon$ prescription such that the denominator in the rule d.1 becomes of 
the form $E_{inc}-E_{cut}+i\epsilon$, and replaces the factor $(-1)^N$ by $i$ 
in rule d.3, then at $T=0$ the rules precisely reproduce the Green's functions 
of the real time theory. 
This $i\epsilon$ prescription provides a definite analytic continuation of the 
Euclidean result into Minkowski space. 
We are presently investigating whether this continuation is indeed the 
real-time theory at finite temperature as well.
 
\bigskip
\noindent{\Large \bf Acknowledgment}
\medskip
\ret
This work was supported by CONICYT under Grant Fondecyt-1931121.

\newpage
\noindent{\Large \bf Figure captions}
\medskip
\ret
FIG.~1. The lowest order correction to the inverse propagator in the
scalar $g\phi^3$ theory. (a) and (b) are the two possible time-ordered
graphs at $T=0$. (c) to (f) are the thermal contributions derived from
(a) and (b) by snipping the internal line with energy $E_1$: 
in (c) and (e) the thermal legs are {\em internal}; in (d) 
and (f) they are {\em external}. 
We read off $E_{inc}$ as $iP_0$ for (c) and (e), and $iP_0 + E_1$ for (d) and (f). 
Similarly, $E_{cut}$ is $E_1+E_2$ for (c), $E_2$ for (d), $2iP_0+E_1+E_2$ 
for (e), and $2iP_0+E_2$ for (f).

\medskip
\ret
FIG.~2 A 2-loop contribution in $\lambda\phi^4$.  (a) and (b) are the two 
possible time-ordered graphs at $T=0$. (c) and (d) are some of the thermal diagrams 
generated from (a) and (b) by snipping one line: in (c) the thermal legs are 
external; in (d) they are internal. (e) and (f) are some of the diagrams 
generated from (a) and (b) by snipping two lines: in (e) both thermal pairs
are external; in (f) pair 1 is internal and pair 3 external. 

\medskip
\ret
FIG.~3. One-loop contribution to the 3-point function in $g\phi^3$ theory.
(a) and (b) are two possible time orderings at $T=0$. (c) and (d) are some of the
thermal diagrams generated from (a) and (b) by snipping one line: 
in (c) the thermal legs are external; in (d) they are internal. 

\medskip
\ret
FIG.~4. (a) t-channel and (b) s-channel Feynman diagrams for the
$T=0$ forward scattering process
$\phi(P^0,{\bf p})+\phi(-iE_1, {\bf k})\rightarrow
\phi(P^0,{\bf p})+\phi(-iE_1, {\bf k})$.

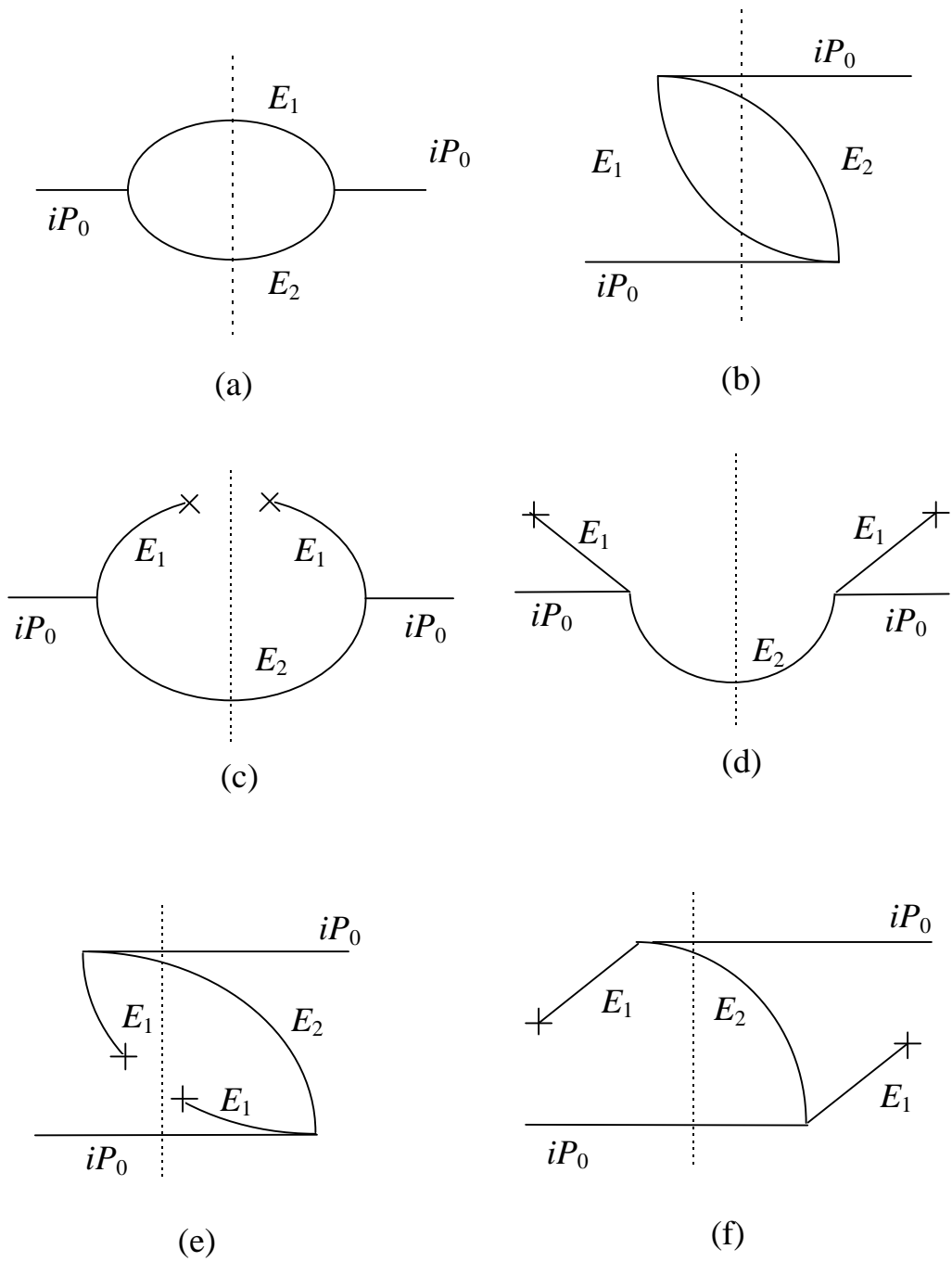
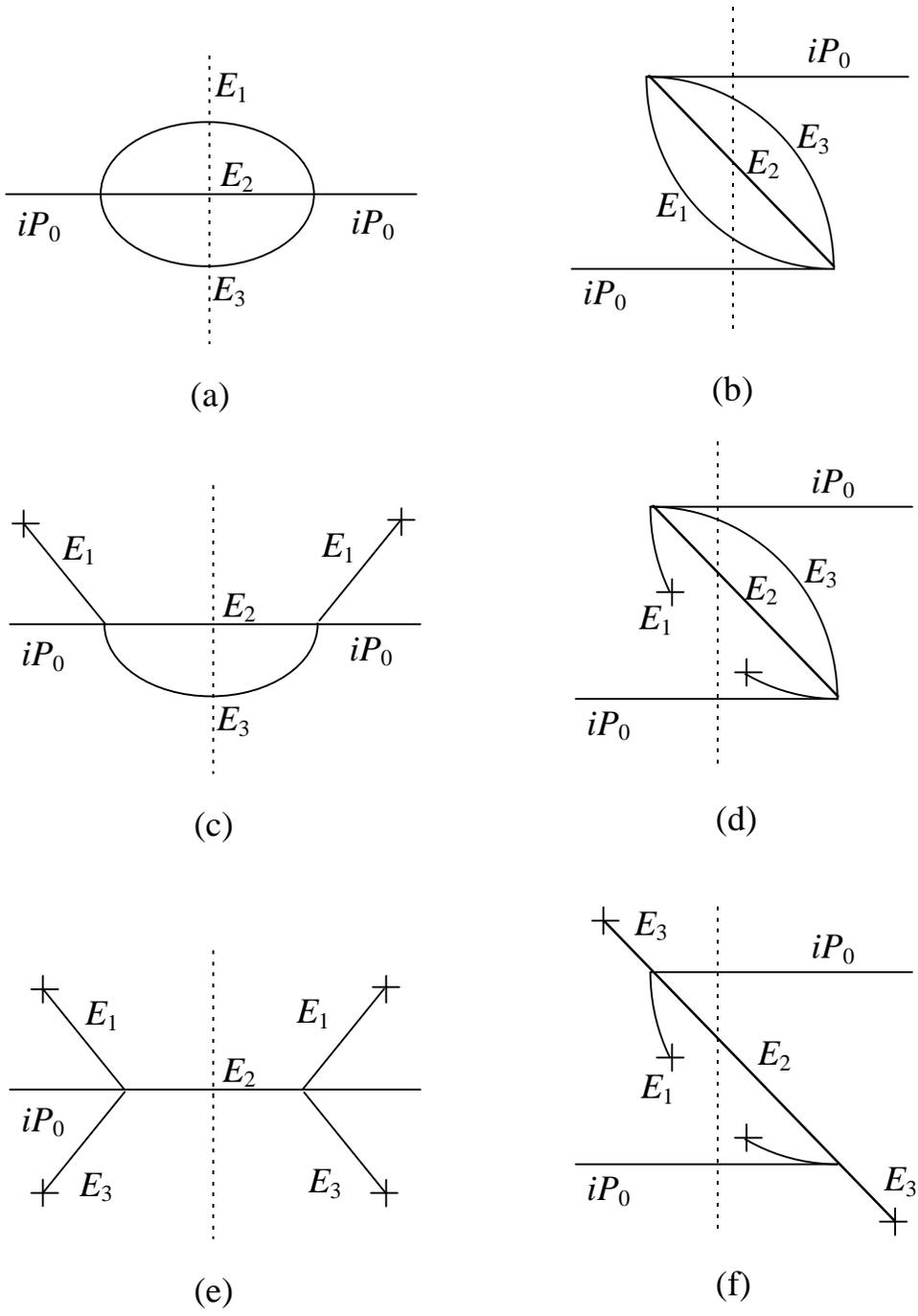
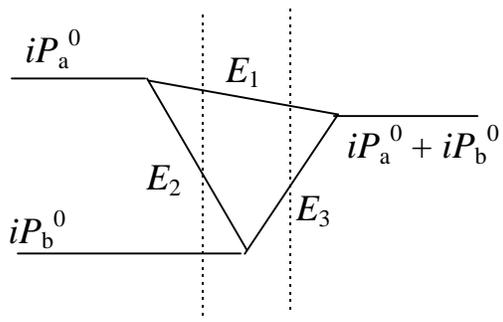
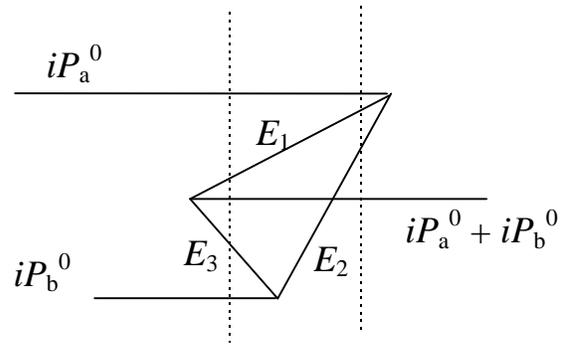
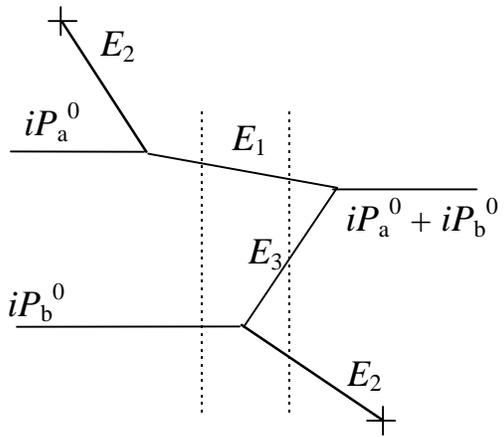
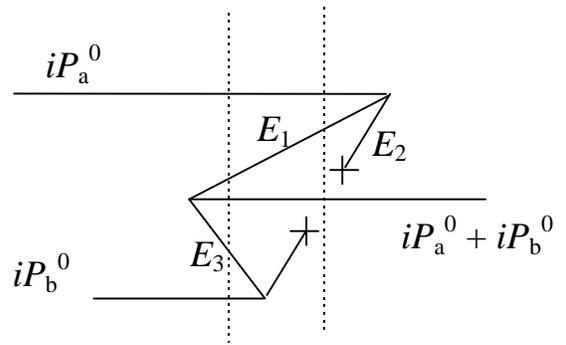
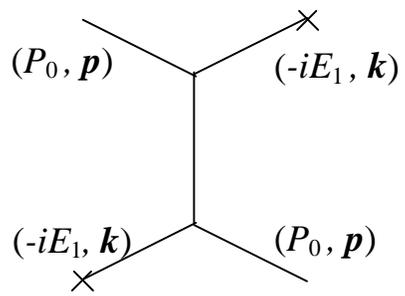
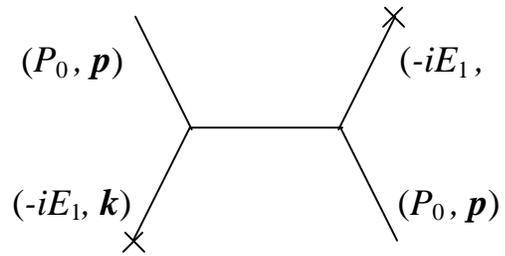
\end{document}